 \documentclass[final,onecolumn,3p,times,sort&compress,12pt]{elsarticle}

\usepackage{graphicx}
\usepackage{dcolumn}
\usepackage{bm}
\usepackage{epstopdf}
\usepackage{bm,hyperref,color,breakurl}

	\usepackage{fancyheadings}
\renewcommand{\thispagestyle}[1]{} 

\begin{document}

\begin{frontmatter}



\title{Hubbard pair cluster in the external fields. Studies of the magnetic properties}


\author{T. Balcerzak}
\ead{t\_balcerzak@uni.lodz.pl}
\author{K. Sza{\l}owski\corref{cor1}}
\ead{kszalowski@uni.lodz.pl}
\address{Department of Solid State Physics, Faculty of Physics and Applied Informatics,\\
University of \L\'{o}d\'{z}, ulica Pomorska 149/153, 90-236 \L\'{o}d\'{z}, Poland}

 \cortext[cor1]{Corresponding author. E−mail address:
kszalowski@uni.lodz.pl}

\date{\today}

\begin{abstract}
The magnetic properties of the two-site Hubbard cluster (dimer or pair), embedded in the external electric and magnetic fields and treated as the open system, are studied by means of the exact diagonalization of the Hamiltonian. The formalism of the grand canonical ensemble is adopted. The phase diagrams, on-site magnetization, spin-spin correlations, mean occupation numbers and hopping energy are investigated and illustrated in figures. An influence of temperature, mean electron concentration, Coulomb $U$ parameter and external fields on the quantities of interest is presented and discussed.  
In particular, the anomalous behaviour of the magnetization and correlation function vs. temperature near the critical magnetic field is found. Also, the effect of magnetization switching by the external fields is demonstrated.

\end{abstract}


\end{frontmatter}

\section{Introduction}

The Hubbard model, since its formulation \cite{Anderson, Hubbard, Gutzwiller, Kanamori}, has been intensively studied within many areas of solid state physics \cite{Chen, Ho, Hirsch3, Robaszkiewicz1, Robaszkiewicz2, Hirsch, Lieb, Lieb3,  Hirsch2, Sorella, Pelizzola, Janis, Staudt, Peres, Kent, Zaleski, Schumann, Schumann2, Cisarova, Cencarikova, Galisova, Galisova2, Harris, Silantev, Hasegawa,Hasegawa2,Spalek, Longhi, Kozlov, Juliano, Joura, Li, Alvarez, McKenzie, Fuchs, Rohringer, Kozik, Karchev, Yamada, Claveau, Lieb2, Shastry, Su, Mancini, Tocchio, Dang, Mermin, Nolting, Dombrowsky, Feldner, Szalowski, Barnas, Chao, Yosida, Tasaki, Micnas, Georges, Hirschmeier,Mielke,B-Sz, Szalowski-Acta, Wortis,Wortis2,Wortis3}. Its applications concern such problems as the description of metal-insulator transition, magnetism of itinerant electrons, studies of high-temperature superconductors, optical lattices or magnetism of graphene nanostructures. 

It should be noted that the exact results for this model for infinite systems are restricted merely to several particular cases like, for instance, the solution in 1D case \cite{Lieb2, Shastry,Su,Mancini}, Mermin-Wagner theorem for 2D systems \cite{Mermin, Nolting, Tasaki}, or Lieb theorems for the ground state \cite{Lieb}. Therefore, many efforts have been undertaken to apply the model to small atomic clusters \cite{Schumann,Schumann2, Cisarova,Cencarikova,Galisova,Galisova2,Harris,Silantev,Hasegawa,Hasegawa2,Spalek,Longhi,Juliano,Kozlov,Li,Joura,Alvarez,McKenzie,Amendola,Iglesias,Acquarone, B-Sz, Szalowski-Acta,Wortis}, where the exact solution can be obtained either by analytical methods or by computer analysis. Such a direction of theoretical studies coincides with the recent trends in experimental solid state physics, where special attention is drawn to the application of nanoclusters and nanotechnology.

It is quite clear that the properties of nanoclusters are influenced by the interaction of these systems with the environment. From the thermodynamic point of view such interaction may involve  the coupling with the external fields, flow of heat or flow of mass. Therefore, the methods of theoretical description for the most general case, when all these interactions are present, should be based on the grand canonical ensemble. 

Taking the above facts into account, in our previous paper \cite{B-Sz} we developed an exact analytical method for studies of the Hubbard pair-cluster embedded simultaneously in two fields: magnetic and electric one, and exchanging the electrons with its environment. In that general approach, the mean number of electrons localized on the cluster can be a non-integer number at a finite temperature, whereas the two fields can compete with each other, influencing the cluster properties. Considering such a cluster (dimer), the exact analytical diagonalization of the Hamiltonian has been performed and the grand  partition function has been found.

In the previous paper \cite{B-Sz} we concentrated only on the calculations of the chemical potential, which is a necessary parameter for the full thermodynamic description of the open system. Those comprehensive investigations paved the way for further statistical-thermodynamic studies of the system in question, including the analysis of the magnetic properties.

The aim of the present paper is to study the magnetic properties of the Hubbard pair-cluster, embedded simultaneously in the magnetic and electric fields and exchanging the electrons with the cluster environment, whereas the system is in thermal equilibrium. The theoretical method, developed in \cite{B-Sz}, will be used here for the studies of the magnetic phase diagrams, on-site magnetizations, spin-spin correlation function, mean hopping energy and mean occupation numbers. The influence of the external fields, temperature, as well as the variable electron concentration on these quantities will be presented.

The paper is organized as follows:  In the next Section the theoretical model will be shortly outlined. In the successive parts of the paper the numerical results will be presented in figures and discussed. Finally, in the last section, the main results will be summarized and the conclusions will be drawn.

\section{Theoretical model}

The Hubbard Hamiltonian for a pair of atoms $(a,b)$ embedded in the external fields  is given in the form of:
\begin {eqnarray}
\mathcal{H}_{a,b}&=&-t\sum_{\sigma=\uparrow,\downarrow}\left( c_{a,\sigma}^+c_{b,\sigma}+c_{b,\sigma}^+c_{a,\sigma} \right)+U\left(n_{a,\uparrow}n_{a,\downarrow}+n_{b,\uparrow}n_{b,\downarrow}\right)\nonumber\\
&&-H\left(S_a^z+S_b^z\right) -V\left(n_{a}-n_{b}\right),
\label{eq1}
\end {eqnarray}
where $t>0$ is the hopping integral and $U\ge0$ is on-site Coulomb repulsion energy.  The term with $H=-g\mu_{\rm B}H^z$ introduces an external magnetic field $H^z$, while $V=E|e|d/2$ denotes the electric potential, applied to $a$ and $b$ atoms and resulting from the presence of an external electric field $E$ oriented along the pair, with $d$ being the interatomic distance, whereas $e$ is the electron charge. In the present model we assume that the hopping integral is a constant parameter, independent on the external fields. 
In Hamiltonian (\ref{eq1}) $c_{\gamma,\sigma}^+$ and $c_{\gamma,\sigma}$ are the electron creation and annihilation operators, respectively, and $\sigma=\downarrow,\uparrow$ denotes the spin state.  Total occupation number operators $n_{\gamma}$ for site $\gamma = a,b$, are defined as a sum of occupation operators for given spin $n_{\gamma,\sigma}$ which, in turn, are expressed by the product of creation and annihilation operators: $n_{\gamma}=\sum_{\sigma}n_{\gamma,\sigma}=\sum_{\sigma}c_{\gamma,\sigma}^+c_{\gamma,\sigma}$. The resulting $z$-component of the spin on a given atom, $S_{\gamma}^z$, is defined as $S_{\gamma}^z=\left(n_{\gamma,\uparrow}-n_{\gamma,\downarrow}\right)/2$.

Treating the pair-cluster as an open electronic system, the exact analytic diagonalization of the Hamiltonian has been performed in Ref. \cite{B-Sz} and the grand partition function has been found there. Then, the grand thermodynamic potential $\Omega_{a,b}$ has been obtained in the form of:
\begin {equation}
\Omega_{a,b}=-k_{\rm B}T \ln \mathcal{Z}_{a,b}=-k_{\rm B}T \ln \{ {\rm Tr}_{a,b} \,\exp \lbrack -\beta \left(\mathcal{H}_{a,b}-\mu\left(n_a+n_b\right)\right)\rbrack \},
\label{eq2}
\end {equation}
where $\mathcal{Z}_{a,b}$ is the grand partition function and $\mu$ is the chemical potential.

 The knowledge of $\Omega_{a,b}$ enables the self-consistent calculations of all the thermodynamic properties of the cluster in question provided the chemical potential is determined. The chemical potential $\mu$ of the electrons fulfils the relationship:
\begin {equation}
\left(\frac{\partial \Omega}{\partial \mu}\right)_{T,H,E}=-\left(\left<n_a\right>+\left<n_b\right>\right)
\label{eq3}
\end {equation}
where $\left<n_a\right>$ and $\left<n_b\right>$ are the thermodynamic mean values of the total occupation number operators for $\gamma =a,b$ sites, respectively. The partial derivative in Eq.(\ref{eq3}) is calculated at constant temperature $T$ and external fields $H$ and $E$. The averages of occupation number operators in Eq. (\ref{eq3})  can be calculated from the formula:

\begin {equation}
\left<n_{\gamma} \right>={\rm Tr}_{a,b} \lbrack n_{\gamma}\; \rho_{a,b} \rbrack,
\label{eq4}
\end {equation}
where $\rho_{a,b}$ is the statistical operator for the grand canonical ensemble:
\begin {equation}
\rho_{a,b}=\frac{1}{\mathcal{Z}_{a,b}}\, \exp \lbrack -\beta \left(\mathcal{H}_{a,b}-\mu\left(n_a+n_b\right)\right)\rbrack.
\label{eq5}
\end {equation}
In order to find $\mu$ it is convenient to define the parameter $x$ denoting the mean number of electrons per lattice site, i.e., the electron concentration:
\begin {equation}
x=\left(\left<n_a\right>+\left<n_b\right>\right)/2.
\label{eq6}
\end {equation}
Then, using the relationships  (\ref{eq6}) and (\ref{eq4}) the chemical potential $\mu$ can be self-consistently found as a function of $T$, $H$, $E$, $U$ and $x$. We note that for the Hubbard  pair-cluster, $x$ can take the values from the interval $[0,2]$.

The extensive calculations of the chemical potential have been performed
 in Ref. \cite{B-Sz} and they form a basis for the present studies of magnetic properties. After obtaining $\mu$, the statistical operator $\rho_{a,b}$ is at our disposal and the thermodynamic mean value of any operator $O$ can be calculated as:
\begin {equation}
\left< O\right>={\rm Tr}_{a,b} \lbrack O \, \rho_{a,b} \rbrack.
\label{eq7}
\end {equation}
In this way, based on the exact diagonalization of the Hamiltonian,  we can calculate the average values of the operators essential for the present study, which give, among other things, the local magnetization $\left< S_{\gamma}^z \right>$, magnetic correlation function $\left< S_{a}^z S_{b}^z \right>$, hopping energy $-t\sum_{\sigma=\uparrow,\downarrow}\left< c_{a,\sigma}^+c_{b,\sigma}+c_{b,\sigma}^+c_{a,\sigma} \right>$ and so on.
In the case of open system, these quantities are parametrized by the arbitrary electron concentration $x$, temperature $T$, magnetic field $H$, electric field $E$, and Coulomb $U$-parameter.
The results of numerical calculations  performed in the framework of the above formalism are presented in the next Section. 

\section{Numerical results and discussion}

\begin{figure}[h]
\begin{center}
\includegraphics[width=128mm]{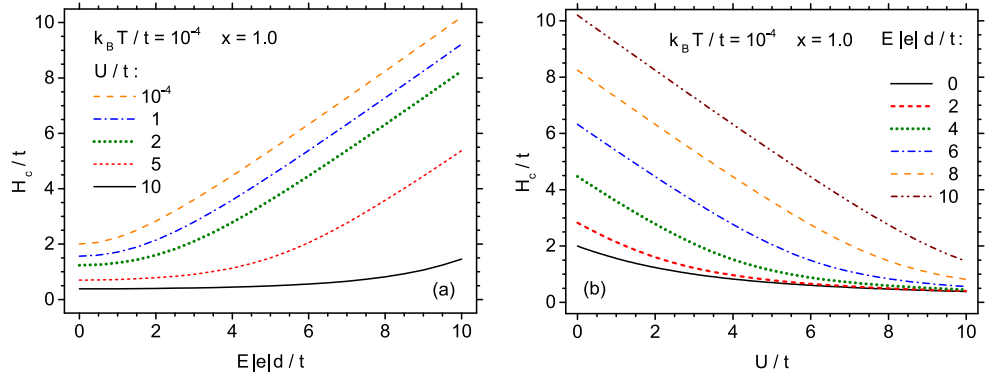}
\vspace{2mm}
\caption{\label{Fig1} The phase diagram presenting the critical magnetic field $H_c/t$: (a) vs. normalized electric field $E|e|d/t$ for several $U/t$-parameters, and (b) vs. Coulomb $U/t$-parameter for several $E|e|d/t$ electric fields. The temperature $k_{\rm B}T/t$=0.0001 and the electron concentration  $x=1$ are assumed.} 
\end{center}
\end{figure}

\begin{figure}[h]
\begin{center}
\includegraphics[width=128mm]{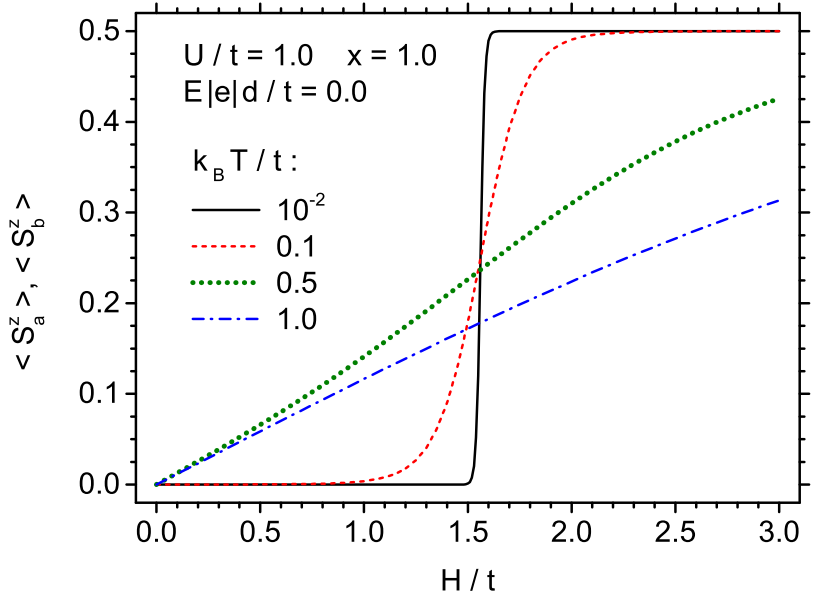}
\vspace{2mm}
\caption{\label{Fig2} The on-site magnetizations, $m_a=\left<S_a^z\right>=m_b=\left<S_b^z\right>$, as the functions of magnetic field $H/t$, for several reduced temperatures $k_{\rm B}T/t$. The $U/t$-parameter is equal to $U/t=1$, the electron concentration corresponds to half-filling, $x=1$, and the electric field $E$ is absent. }
\end{center}
\end{figure}

\begin{figure}[h]
\begin{center}
\includegraphics[width=128mm]{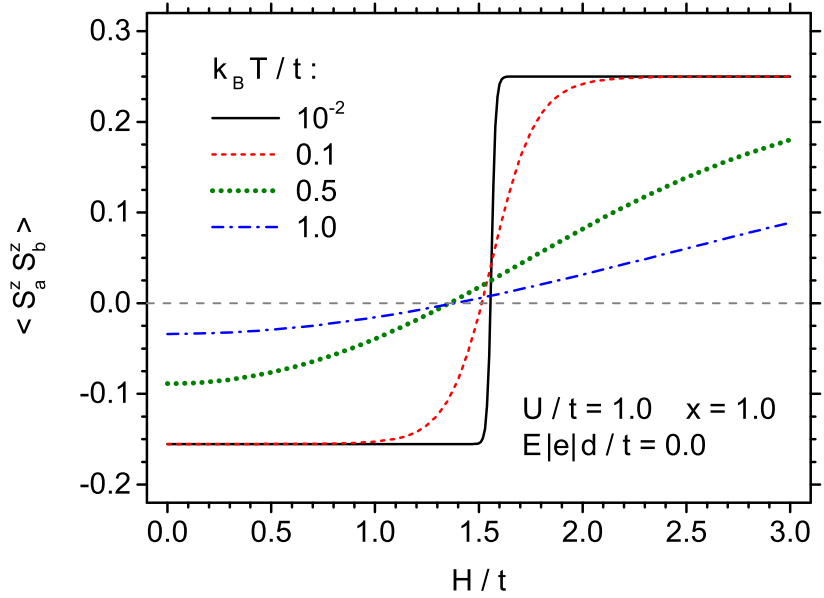}
\vspace{2mm}
\caption{\label{Fig3} The magnetic correlation function $\left<S_a^z S_b^z\right>$ vs. magnetic field $H/t$ for several reduced temperatures $k_{\rm B}T/t$. The remaining parameters are the same as in Fig.~\ref{Fig2} .}
\end{center}
\end{figure}

\begin{figure}[h]
\begin{center}
\includegraphics[width=128mm]{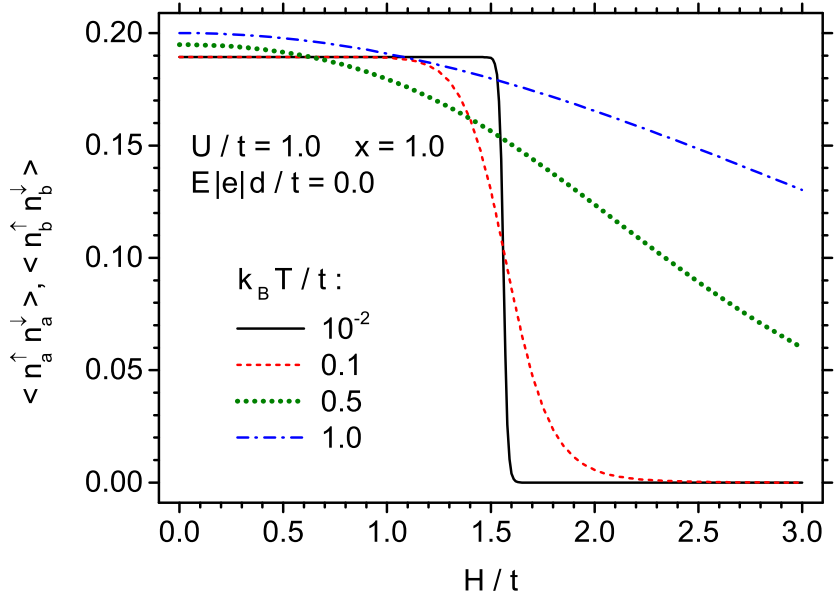}
\vspace{2mm}
\caption{\label{Fig4} The on-site mean occupation number correlations, $\left<n_{a,\uparrow} n_{a,\downarrow}\right>=\left<n_{b,\uparrow} n_{b,\downarrow}\right>$, as the function of magnetic field $H/t$, for several reduced temperatures $k_{\rm B}T/t$. The remaining parameters are the same as in Figs.~\ref{Fig2}  and ~\ref{Fig3} .}
\end{center}
\end{figure}

\begin{figure}[h]
\begin{center}
\includegraphics[width=128mm]{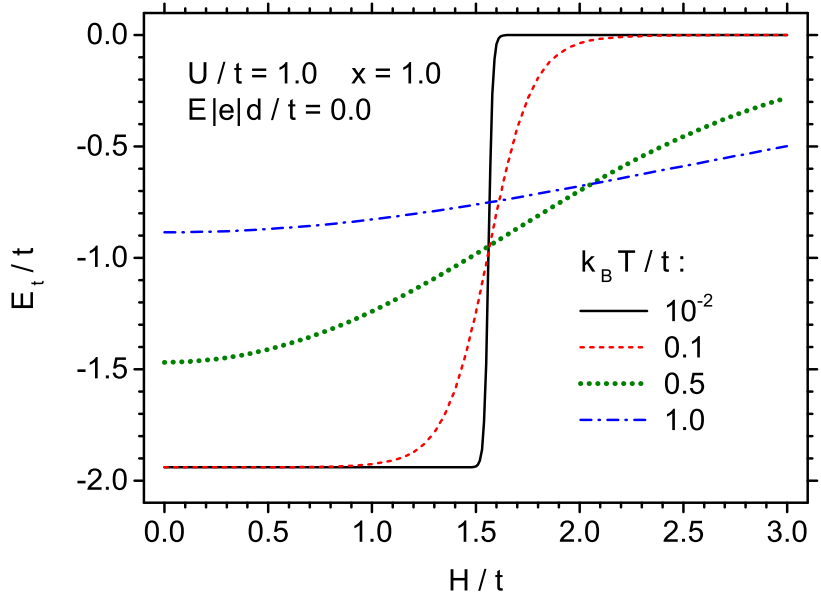}
\vspace{2mm}
\caption{\label{Fig5} Mean hopping energy $E_t/t= -\sum_{\sigma=\uparrow,\downarrow}\left< c_{a,\sigma}^+c_{b,\sigma}+c_{b,\sigma}^+c_{a,\sigma} \right>$ as the function of magnetic field $H/t$, for several reduced temperatures $k_{\rm B}T/t$.
The remaining parameters are the same as in Figs.~\ref{Fig2} - ~\ref{Fig4} .}
\end{center}
\end{figure}

\begin{figure}[h]
\begin{center}
\includegraphics[width=128mm]{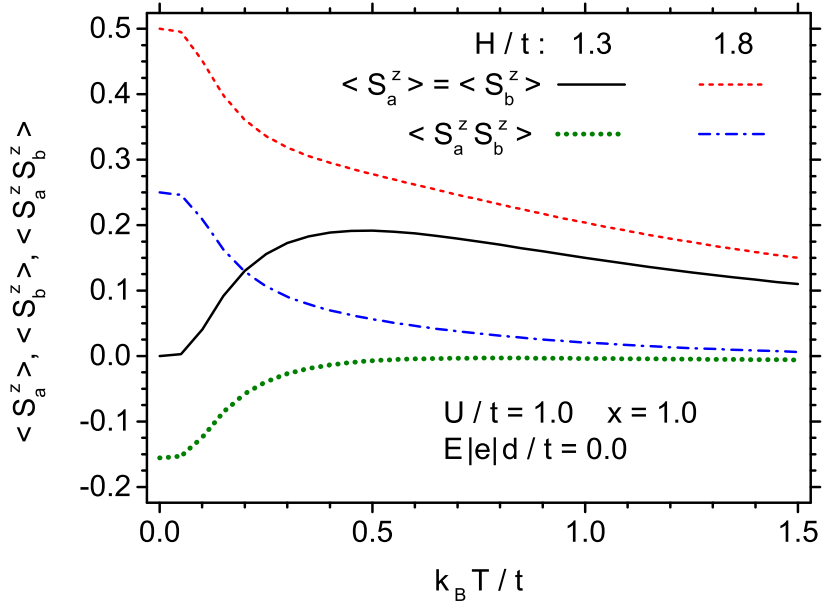}
\vspace{2mm}
\caption{\label{Fig6} On site magnetizations, $m_a=\left<S_a^z\right>=m_b=\left<S_b^z\right>$, as well as the magnetic correlation function, $\left<S_a^z S_b^z\right>$, vs. reduced temperature $k_{\rm B}T/t$, for two magnetic fields: $H/t=1.3$ and $H/t=1.8$. The remaining parameters are: $U/t=1$, $x=1$ and $E=0$. }
\end{center}
\end{figure}

\begin{figure}[h]
\begin{center}
\includegraphics[width=128mm]{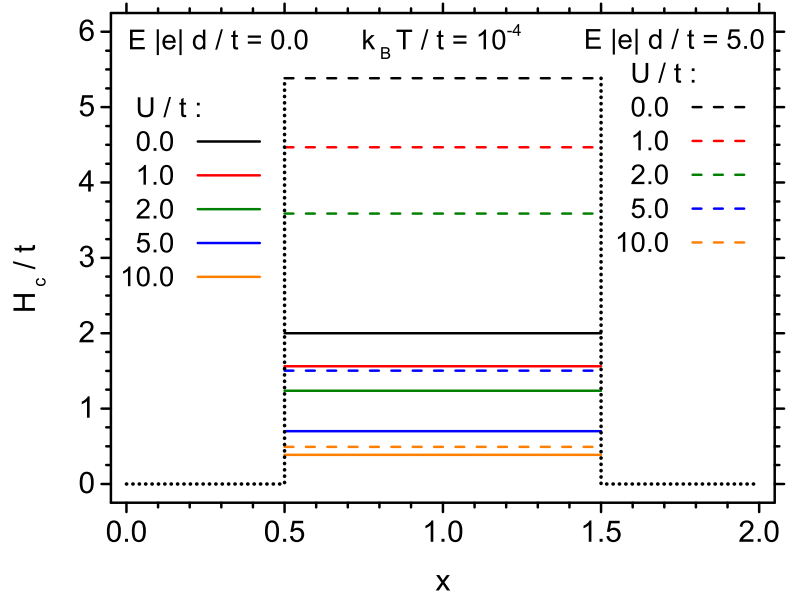}
\vspace{2mm}
\caption{\label{Fig7} The phase diagram showing the critical magnetic field $H_c/t$ vs. electron concentration $x$, for the temperature very close to zero, $k_{\rm B}T/t=0.0001$, when the system is almost in the ground state. Different plots correspond to several $U/t$-parameters and two values of the reduced electric field: $E|e|d/t=0$ and $E|e|d/t=5$.}
\end{center}
\end{figure}

\begin{figure}[h]
\begin{center}
\includegraphics[width=128mm]{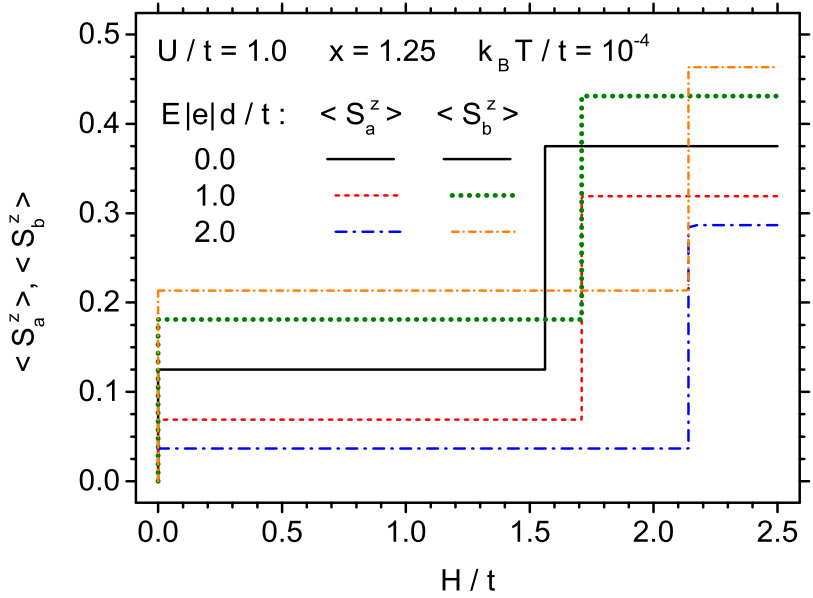}
\vspace{2mm}
\caption{\label{Fig8} The on site magnetizations, $m_a=\left<S_a^z\right>$  and $m_b=\left<S_b^z\right>$, as the functions of the magnetic field $H/t$. The remaining parameters are: Coulomb repulsion parameter is $U/t=1$, electron concentration - $x=1.25$ and temperature $k_{\rm B}T/t=0.0001$. Different plots correspond to various values of the normalized electric field $E|e|d/t$.}
\end{center}
\end{figure}

\begin{figure}[h]
\begin{center}
\includegraphics[width=128mm]{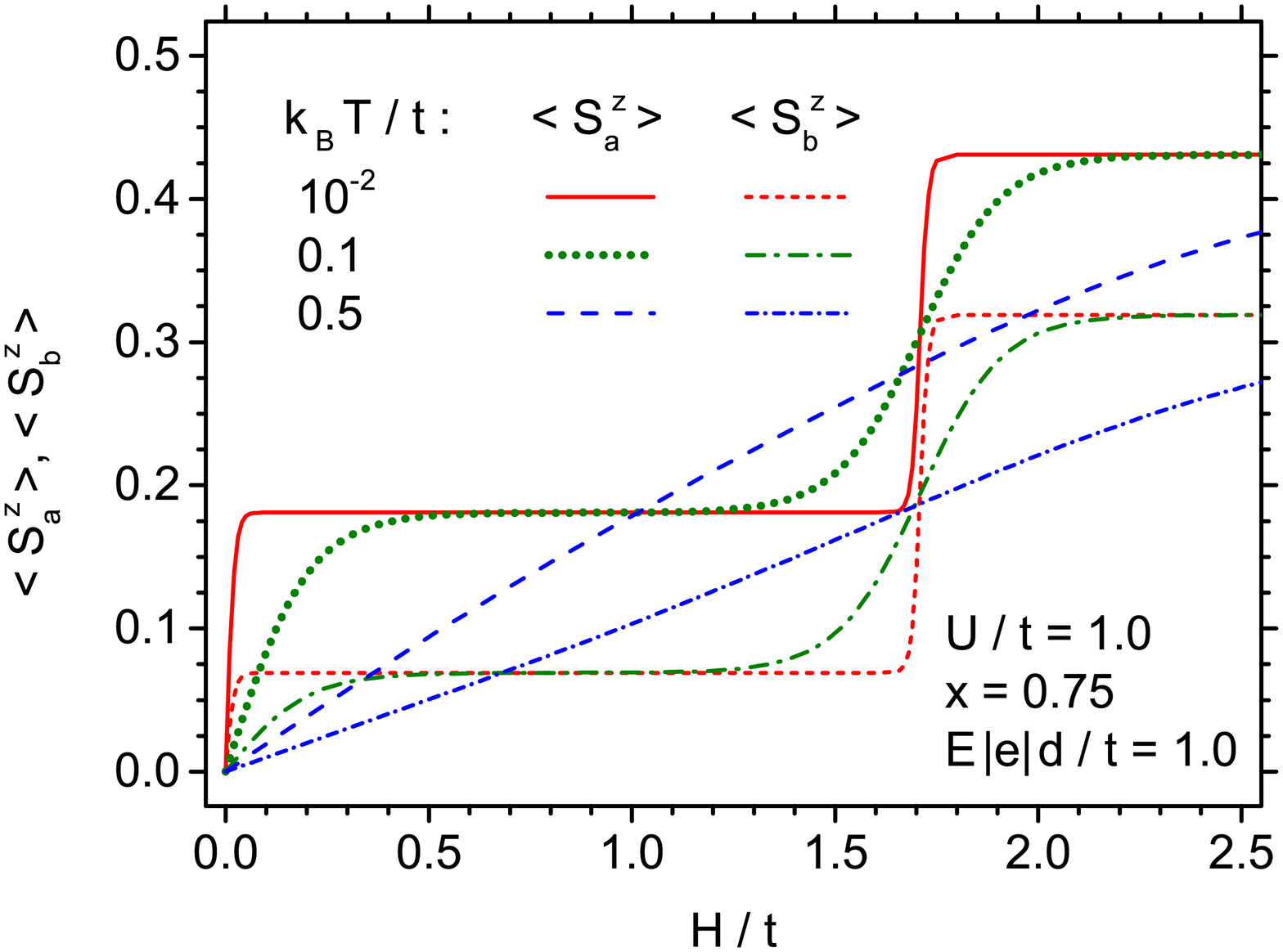}
\vspace{2mm}
\caption{\label{Fig9} The on site magnetizations, $m_a=\left<S_a^z\right>$  and $m_b=\left<S_b^z\right>$, as the functions of the magnetic field $H/t$ for several reduced temperatures $k_{\rm B}T/t$. The remaining parameters are: $U/t=1$, $x=0.75$ and  $E|e|d/t=1$.}
\end{center}
\end{figure}

\begin{figure}[h]
\begin{center}
\includegraphics[width=128mm]{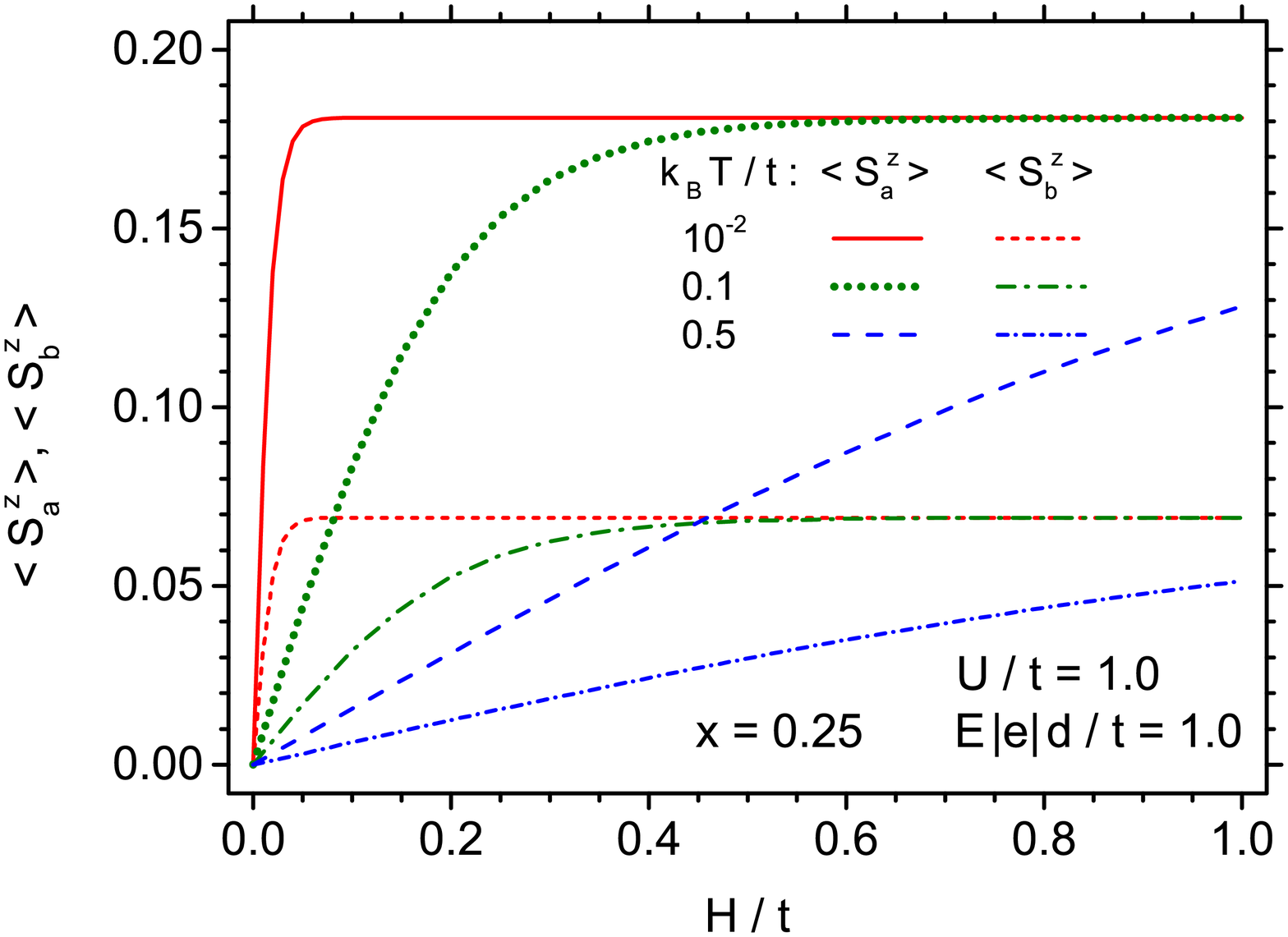}
\vspace{2mm}
\caption{\label{Fig10} The on site magnetizations, $m_a=\left<S_a^z\right>$  and $m_b=\left<S_b^z\right>$, as the functions of the magnetic field $H/t$ for several reduced temperatures $k_{\rm B}T/t$. The remaining parameters are: $U/t=1$, $x=0.25$ and  $E|e|d/t=1$.}
\end{center}
\end{figure}

\begin{figure}[h]
\begin{center}
\includegraphics[width=128mm]{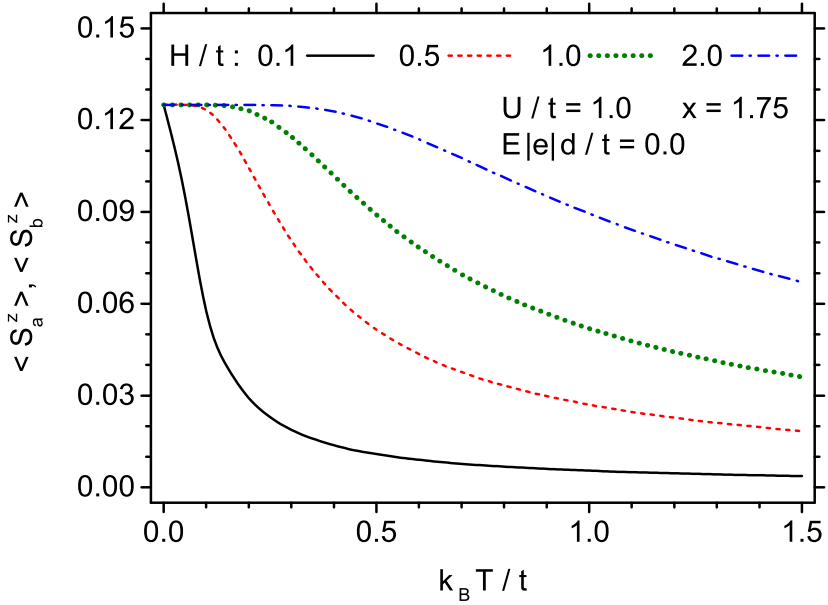}
\vspace{2mm}
\caption{\label{Fig11} The on-site magnetizations,  $m_a=\left<S_a^z\right>=m_b=\left< S_b^z\right>$, vs. dimensionless temperature $k_{\rm B}T/t$, for several values of the external magnetic field $H/t$. The rest of parameters are: $U/t=1$, $x=1.75$ and  $E|e|d/t=0$.}
\end{center}
\end{figure}

\begin{figure}[h]
\begin{center}
\includegraphics[width=128mm]{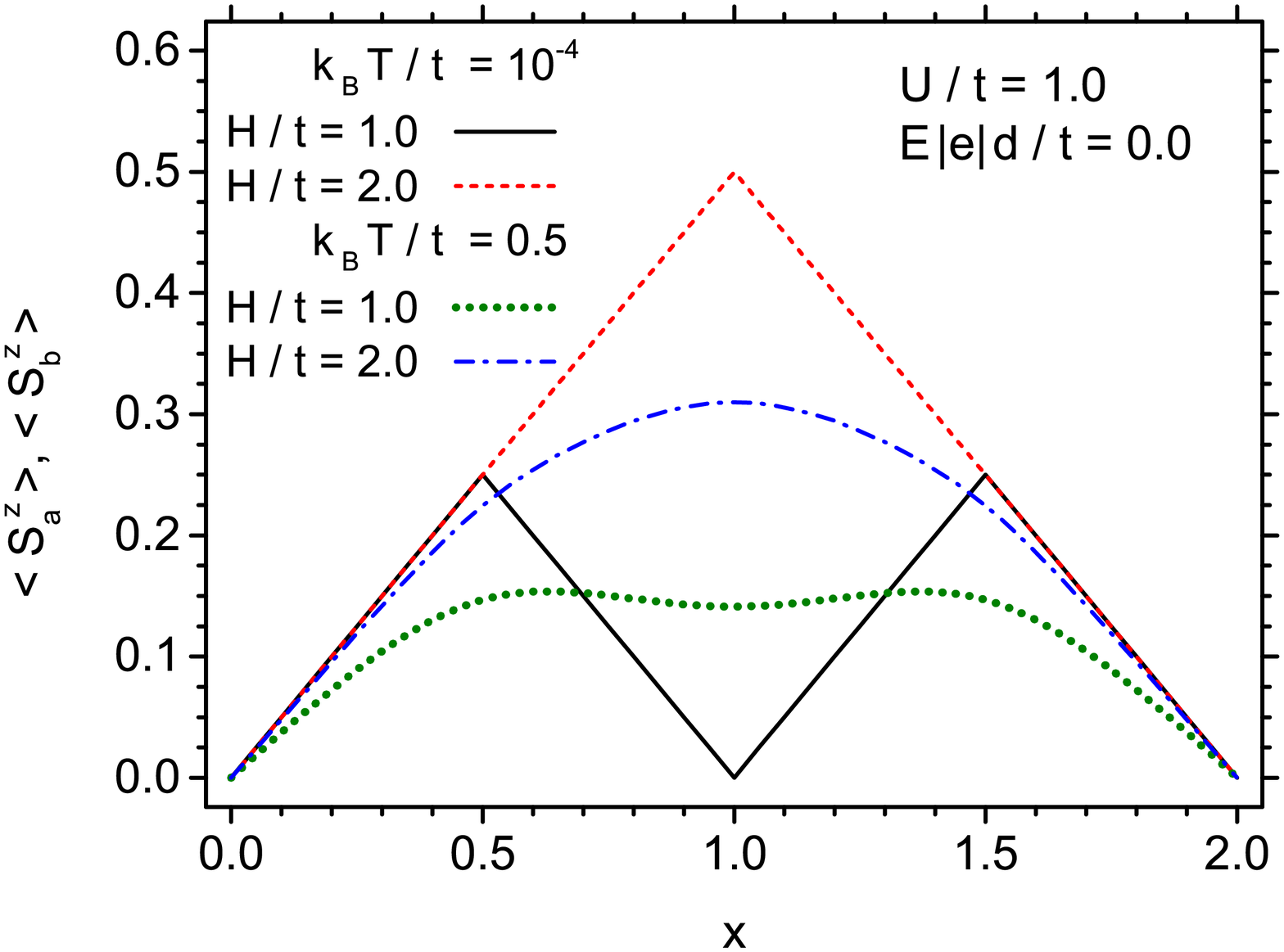}
\vspace{2mm}
\caption{\label{Fig12} The on-site magnetizations,  $m_a=\left<S_a^z\right>=m_b=\left< S_b^z\right>$, vs. electron concentration $x$. Two temperatures: $k_{\rm B}T/t=0.0001$ and $k_{\rm B}T/t=0.5$, as well as two magnetic fields: $H/t=1.0$ and $H/t=2$ are selected. The Coulomb $U/t$-parameter is $U/t=1$ and the electric field $E$ is absent.}
\end{center}
\end{figure}

The numerical results are obtained on the grounds of the model described in the previous Section. They present rigorous calculations of the magnetic properties of the Hubbard pair embedded in the external magnetic and electric fields.

In Fig.~\ref{Fig1} the phase diagram is presented for the temperature very close to  zero, when the system is almost in the ground state ($k_{\rm B}T/t$=0.0001), and for the case of half filling ($x=1$). The curves represent the critical magnetic field $H_c/t$: (a) vs. normalized electric field $E|e|d/t$ for several Coulomb $U/t$ parameters, and (b) as a function of $U/t$ for selected values of $E|e|d/t$. Below the critical field the Hubbard pair is in the paramagnetic state with on-site magnetizations $m_a=m_b=0$. 
The name paramagnetic is used here and throughout the paper in the classical sense, because both the order parameter known from the spin models of antiferromagnets,  $|m_a - m_b|/2$, and the magnetic polarization per atom characterizing ferromagnets, $|m_a + m_b|/2$ are equal to zero. However, at the same time the magnetic pair correlation function, $\left<S_a^z S_b^z\right>$ is negative (as it will be seen in Fig.~\ref{Fig3}), which is characteristic of a short-range antiferromagnetic order. This suggests that also another name like "quantum antiferromagnetic state with zero on-site magnetization" would be adequate.
Above the critical field the ferromagnetically ordered phase with $m_a=m_b=1/2$ occurs. In such ferromagnetic state each electron occupies only one site and takes the maximal spin projection $S^z=1/2$ in the magnetic field direction. In contrast, in the paramagnetic state each electron is delocalized between both atoms, occupying each site with the same probability, and the spins of the electrons are opposite. Thus, the total on-site  magnetization is zero in this state. Higher electric fields require higher critical magnetic fields $H_c$ in order to enforce electron separation between $a$ and $b$ atoms and the spin reversal of one electron (Fig.~\ref{Fig1}(a)). On the other hand, increasing Coulomb $U$-parameter, representing repulsive force between two electrons, facilitates the transition and lowers the critical magnetic field (Fig.~\ref{Fig1}(b)).

The on-site magnetizations $m_a=\left<S_a^z\right>$ and $m_b=\left<S_b^z\right>$ (where $m_a=m_b$) are presented in Fig.~\ref{Fig2} as the functions of magnetic field $H/t$, for several reduced temperatures $k_{\rm B}T/t$. In this case the electric field $E=0$ is assumed, and remaining parameters are: $U/t=1$ and $x=1$. We see that the magnetization curves are increasing functions vs. magnetic field and their shapes are changing for different temperatures. When the system is near the ground state, for $k_{\rm B}T/t=0.01$, the most dramatic changes occur at the critical field, $H_c/t=1.56155$, when the local magnetization jumps from zero value to the saturation. This is in agreement with the discussion of Fig.~\ref{Fig1}. With increasing temperature the magnetic transition becomes diffused and the curves take increasingly linear shape. It can be noticed that for a fixed magnetic field, lower than the critical value, the magnetization is at first an increasing function of temperature and then starts to decrease. On the other hand, for $H>H_c$, the magnetization is always a decreasing function when temperature increases. Such an anomalous behaviour of the magnetization vs. temperature, for $H<H_c$, will be further discussed in Fig.~\ref{Fig6}.

In Fig.~\ref{Fig3} the magnetic pair correlation function, $\left<S_a^z S_b^z\right>$, is shown vs. magnetic field, for different temperatures. All remaining parameters are the same as in Fig.~\ref{Fig2}. The behaviour of magnetic correlation is qualitatively similar to the behaviour of magnetization. We found that when the system is near the ground state the correlation function jumps from the value of -0.1553 in paramagnetic phase to the saturation value of 0.25 in the ferromagnetic phase. In general, correlations change the sign when the magnetic field is strong enough, and in the ground state the most rapid change takes place at $H_c$. For $H \to 0$ the correlations are negative and they increase when temperature increases, whereas for $H>H_c$ they are positive and show an opposite tendency. In some small range of $H<H_c$ an anomalous behaviour can be predicted, when correlations are non-monotonic functions of temperature. Such a behaviour is similar to the magnetization dependence discussed in the previous figure, although here the effect is much weaker.

For the same set of parameters as in Figs.~\ref{Fig2} and ~\ref{Fig3}, the on-site occupation correlations, $\left<n_{\alpha,\uparrow} n_{\alpha,\downarrow}\right>$ ($\alpha=a,b$), are presented in Fig.~\ref{Fig4}. We see that the electron occupation correlations are strongest for $H \to0$ and they decrease monotonically with the increase of $H$. As before, the most rapid changes take place near the ground state at the critical field $H_c$, and for $H>H_c$ the correlations drop to zero. A lack of correlations corresponds to the spatial separation of two electrons when they are localized on $a$ and $b$ atoms. In general, in $H \to 0$ limit and for $H>H_c$,  the occupation correlations are increasing functions of temperature. In some region of $H<H_c$, an anomalous behaviour can be seen, when at first the correlations are decreasing and then start to increase when temperature increases. 

Mean hopping energy $E_t/t= -\sum_{\sigma=\uparrow,\downarrow}\left< c_{a,\sigma}^+c_{b,\sigma}+c_{b,\sigma}^+c_{a,\sigma} \right>$ is presented in Fig.~\ref{Fig5} as a function of magnetic field $H/t$ for different temperatures $k_{\rm B}T/t$. As before, the constant parameters are: $U/t=1$, $x=1$ and the electric field is set to zero. We see that the most rapid changes of the energy occur for $k_{\rm B}T/t=0.01$ curve, in vicinity of the critical field $H_c$. In the ground state, for $H>H_c$, the hopping energy is equal to zero, which is connected with the freezing of the movement of the electrons. For $H<H_c$ the hopping energy is an increasing function of temperature, whereas for  $H\gg H_c$ just the opposite tendency is evident. In some region of $H>H_c$ an anomalous behaviour can be seen, where, for constant field, the hopping energy changes non-monotonically vs. temperature. As before, this effect is caused by flattening of the curves when temperature increases.

On site magnetizations,  $m_\alpha=\left<S_\alpha^z\right>$ ($\alpha=a,b$), and magnetic correlation functions, $\left<S_a^z S_b^z\right>$, are simultaneously shown in Fig.~\ref{Fig6} vs. dimensionless temperature $k_{\rm B}T/t$. To complement the previous figures, the same parameters: $U/t=1$, $x=1$ and $E=0$ are assumed. The curves are presented for two different magnetic fields: $H/t=1.3$ which is lower than the critical field, i.e., $H_c/t=1.56155$, and for $H/t=1.8$ which is above $H_c$. An anomalous behaviour of the magnetization for $H/t=1.3$ is seen as a broad maximum vs. temperature, which confirms the conclusion drawn from Fig.~\ref{Fig2}. Regarding the correlation function, the anomaly is hardly seen for $H/t=1.3$, since this field is too far from $H_c$ and the effect itself is less spectacular. For $H/t=1.8$, both the magnetization and the correlation are decreasing functions of temperature and correlations are positive (ferromagnetic). In the ground state, the saturation values are: $m_\alpha=0.5$ and $\left<S_a^z S_b^z\right>=0.25$.

In Fig.~\ref{Fig7} the phase diagram, illustrating the critical field, $H_c/t$, above which the phase transition to saturated state takes place, is presented as a function of the mean number of electrons per atom, $x$. The temperature is very close to zero, $k_{\rm B}T/t=0.0001$, and two electric fields: $E|e|d/t=0$ and $E|e|d/t=5$ are selected for comparison. Different lines  correspond to various Coulomb repulsive parameters $U/t$. The saturated phase in the ground state is characterized by the following rule for total pair magnetization: $\left<S_a^z+S_b^z\right>=x$  for $0\leq x\leq 1$, and $\left<S_a^z+S_b^z\right>=2-x$ for $1\leq x\leq 2$. Thus, for $x=1$, the saturation magnetization amounts to $\left<S_a^z+S_b^z\right>=1$, in agreement with the previous figures. It is seen that the diagram is symmetric with respect to $x=1$, which corresponds to the electron-hole symmetry, and in the regions  $0<x<0.5$ and $1.5<x<2$ the critical field is zero. It means that in these regions already the infinitesimal magnetic field brings the electron system to the ordered state, irrespective of $U$ and $E$. In the interval $0.5<x<1.5$, excluding the point $x=1$, the system for $E=0$ is in a non-saturated ferromagnetic state below the critical field $H_c$, and above $H_c$ the jump of magnetization to the saturation value in the ferromagnetic phase takes place. In the same interval, for $E>0$ and $H<H_c$, the system is in a ferrimagnetic state with reduced magnetization, with  $m_a\ne m_b$, whereas for $H>H_c$ the jump of the on-site magnetizations to some stronger ferrimagnetic state occurs. Above $H_c$ the total pair magnetization in the ferrimagnetic state obeys the rule for saturation value given above, while the atomic pair is electrically polarized. At $x=1$, as discussed in previous figures, the system is in paramagnetic state up to the critical field $H_c$, above which the transition to ferromagnetic ordering takes place. In this case the critical fields $H_c$  have been presented in Fig.~\ref{Fig1}. It is remarkable that the critical magnetic fields are constant in the interval $0.5<x<1.5$, irrespective of electron filling $x$, however, they strongly depend on $U$-parameter and electric field $E$. Such dependency has not been found in the remaining intervals $0<x<0.5$ and $1.5<x<2$, where $H_c=0$.

In Fig.~\ref{Fig8} illustration of the on-site magnetizations, $\left<S_a^z\right>$ and $\left< S_b^z\right>$,  vs. magnetic field $H/t$ is shown for $U/t=1$, $x=1.25$ and $k_{\rm B}T/t=0.0001$. Three values of the electric field are chosen: $E|e|d/t=0$, 1 and 2. For $E=0$, $\left<S_a^z\right>=\left< S_b^z\right>$, and when $H<H_c$ the system is in a non-saturated ferromagnetic state. When $H>H_c$ the ferromagnetic saturated phase occurs with $\left<S_a^z+S_b^z\right>=0.75$. For $E>0$ the corresponding phases, occurring below and above $H_c$, are ferrimagnetic with $\left<S_a^z\right>\ne \left< S_b^z\right>$. It can be noted that the splitting of magnetizations is symmetric around the magnetization value for $E=0$ and it increases with the increase of $E$. In particular, the total saturation magnetization of the ferrimagnetic phase is the same as the one of the ferromagnetic phase. These conclusions are in agreement with the results of Fig.~\ref{Fig7}. For $E>0$ the electric polarization of the pair takes place, since the electron occupation numbers $\left<n_a\right>\ne \left< n_b\right>$. Without magnetic field (for $H/t=0$), the system is in paramagnetic state for any electric field $E$. This result is in agreement with Fig.~\ref{Fig1}. Finally, it is worth mentioning that for $x=0.75$ and the same remaining parameters as in Fig.~\ref{Fig8}, the magnetization curves look analogously to these in Fig.~\ref{Fig8}, with the same total saturation value. As mentioned before, this is based on the electron-hole symmetry around $x=1$. The only difference is that $\left<S_a^z\right>$ and $\left< S_b^z\right>$ values are interchanged with respect to those presented for $x=1.25$.

In order to illustrate how the magnetization curves change with temperature, in Fig.~\ref{Fig9} we chose three values of $k_{\rm B}T/t$: 0.01, 0.1 and 0.5, for presentation of $\left<S_\alpha^z\right>$ vs. $H/t$ for $\alpha =a,b$. In this case $x=0.75$, $U/t=1$, and the electric field is $E|e|d/t=1$. It is seen that for the lowest temperature both magnetization curves, $\left<S_a^z\right>$ and $\left< S_b^z\right>$, are step-wise functions with the steps at $H/t$=0 and $H_c/t \approx 1.7$. In higher temperature, for instance, for $k_{\rm B}T/t$= 0.1, the steps become smooth and they finally vanish, as it is already seen for $k_{\rm B}T/t$= 0.5.
For $H/t$=0 the system is in paramagnetic state, with  $\left<S_a^z\right>=\left< S_b^z\right>=0$. On the other hand, if $H/t$ is large enough, both magnetizations tend towards their saturation values, with $\left<S_a^z + S_b^z\right>=0.75$, which is most easily seen for $k_{\rm B}T/t$= 0.01. It should be emphasized that the ferrimagnetic ordering is a consequence of applying the non-zero electric field ($E|e|d/t=1$), and when this field is removed the system comes back to the ferromagnetic state.

In Fig.~\ref{Fig10} we present analogous magnetization curves for lower electron filling, namely, $x=0.25$. All remaining parameters are the same as in Fig.~\ref{Fig9}. In this case, only one magnetization step occurs at $H$=0, which is seen for $k_{\rm B}T/t$= 0.01. It means that the critical magnetic field, for which the system takes the ordered (ferrimagnetic, saturated) ground state, is equal to $H_c/t$=0. This conclusion is in agreement with the phase diagram presented in Fig.~\ref{Fig7} for arbitrary $x$. Again, when the temperature increases, the magnetization step becomes diffused and it eventually vanishes if the temperature is large enough. Another important fact is that, for fixed $H$, an increase in temperature leads always to the decrease in magnetization. Thus, for $x=0.25$, the anomalous behaviour of magnetization discussed previously does not occur. The saturation magnetization in this case is low and equal to $\left<S_a^z+S_b^z\right>=x=0.25$, as discussed in Fig.~\ref{Fig7}. It should be mentioned again that, owing to the symmetry around $x=1$, the identical magnetization curves can be obtained for $x=1.75$, the only difference being that magnetizations $m_a=\left<S_a^z\right>$ and $m_b=\left< S_b^z\right>$ would be interchanged.

In Fig.~\ref{Fig11}, for the electron filling parameter $x=1.75$, the magnetization curves are presented  vs. temperature for $U/t=1$. In this case we find the ferromagnetic state, $\left<S_a^z\right>=\left< S_b^z\right>$, since the electric field is set to zero. Different curves correspond to various constant magnetic fields. Since all these fields are lying above the critical field, $H>H_c=0$, the system is in ferromagnetically saturated state at $T=0$, with $\left<S_a^z+S_b^z\right>=2-x=0.25$. An increase in temperature causes the decrease in magnetization tending to the limiting zero value, with no sign of such anomaly which has been discussed in previous figures, for $x=1$ and $H<H_c$. Higher magnetic fields make the decrease in magnetization vs. temperature slower, yet they have no effect on the saturation value at $T=0$. It has been checked that different $U$-parameters have limited influence on the curves presented in this figure. This fact is in agreement with Fig.~\ref{Fig7}, where for $x=1.75$ there is no influence of $U$ on the ground-state phase diagram. 

It is interesting to see how the magnetization depends on the filling parameter $x$. Such dependencies are presented in Fig.~\ref{Fig12} for two very different temperatures and two magnetic fields. The electric field is absent here, $E=0$, in order to avoid ferrimagnetic states, and $U$-parameter is equal to 1. For such $U$-parameter and $x=1$ the value of $H/t=1$ lies below the critical field, $H_c=1.56155$, whereas $H/t=2$ lies above it. It is seen that the most spectacular changes of the magnetic properties for these two fields can be found for $x=1$. We note that the curves presented in this figure are symmetric with respect to $x=1$, and the magnetization is zero for $x=0$ and $x=2$ (when each atom possesses 2 electrons with opposite spins). It can be seen that for $k_{\rm B}T/t$= 0.0001 the magnetization behaves linearly vs. $x$, and in the intervals $0<x<0.5$, as well as $1.5<x<2$, the lines for both magnetic fields are the same. This is because the saturation in these regions is reached for any infinitesimal magnetic field, according to the diagram presented in Fig.~\ref{Fig7}. However, in the interval $0.5<x<1.5$, remarkable differences for both magnetic fields are visible. In particular, for $H/t=2$, the magnetization for $x\le 1$ can be described by the formula $m=x/2$, whereas for $x\ge 1$ the rule $m=1-x/2$ is fulfilled. On the other hand, for $H/t=1$ and $x=1$ the magnetization amounts to zero, since, as discussed previously, the system is in paramagnetic state. As we see, the  magnetization for $x=1$ in the ground state can be rapidly switched between 0 and 1/2 values by applying the magnetic field slightly lower or higher than the critical field $H_c$. Increasing temperature spoils this spectacular switching effect, because the changes of magnetization become smaller and the curves are no more linear. On the basis of this figure it also can be concluded that, for $H/t=2>H_c/t$, an increase in temperature  brings the decrease in magnetization from its saturation value, which is characteristic of given $x$. However, for $H/t=1$ and in the vicinity of $x=1$ we have $H<H_c$ and the opposite effect takes place, i.e., the magnetization increases with temperature (for instance, when comparing the curves for $T>0.0001$ and $T>0.5$). This case corresponds to the anomalous behaviour of magnetization described earlier.

\section{Summary and conclusion}

In the paper, the magnetic properties of the Hubbard pair-cluster have been studied, using the exact diagonalization approach. The cluster has been treated as an open system, exchanging the electrons with its neighbourhood, studied within the full range of average electron concentration $0\le x\le 2$. Moreover, the external magnetic and electric fields have been simultaneously applied to the system, influencing the magnetic properties. The theoretical method has been described in detail in our previous paper \cite{B-Sz}, where the chemical potential, necessary for the present calculations, has already been determined. In the present work we concentrate on the basic magnetic properties like the phase diagrams, on-site magnetizations, mean occupation numbers and the correlation functions.

As a result of numerical calculations the influence of external magnetic and electric fields on the magnetic properties of the Hubbard pair-cluster has been illustrated in figures. The representative parameters of the Hamiltonian and the full range of electron concentration have been considered. Apart from the  ground state, the temperature properties have also been studied.

One of the most interesting findings in the ground state is the critical magnetic field, $H_c$, above which the magnetically ordered, saturated state occurs. At $T=0$ the phase transition is found to be discontinuous; however, it becomes continuous when the temperature increases. The critical magnetic field depends both on the electric field $E$ and parameter $U$, as presented in Fig.~\ref{Fig1}. On the other hand, its dependence on the electron concentration has been illustrated  in Fig.~\ref{Fig7}. The role of electric field consists in a change of the ordered phase from the ferromagnetic (for $E=0$) to the ferrimagnetic one (when $E>0$). The ferrimagnetic state arises from the charge shift between the atoms which are placed in different electric potential and from the influence of $U$-parameter, which enforces the spin reversal of shifted electrons. Resulting ferrimagnetic magnetizations $m_a \ne m_b$ have been illustrated in Figs.~\ref{Fig8}-\ref{Fig10}. 

Regarding the finite temperature calculations, a possibility of anomalous behaviour of both magnetization and (to a smaller extent) the magnetic correlation function has been found in some range of the fields $H<H_c$. Such behaviour has been illustrated in Fig.~\ref{Fig6}, showing a wide maximum of the curves vs. temperature. The origin of this phenomenon lies in flattening of the magnetization (correlation) curves vs. magnetic field, when temperature increases. The effect can be predicted, for instance, on the basis of Fig.~\ref{Fig2}, where a step-wise magnetization representing quantum phase transition for zero temperature becomes an linearly increasing function for higher temperatures. The similar effect can be predicted for the hopping energy in some range of $H>H_c$, as it is seen from Fig.~\ref{Fig5}. The step-wise function in Fig.~\ref{Fig5} corresponds to the fact that in the ground state the kinetic energy is frozen for $H>H_c$. The flattening of the curves when temperature increases, leads in some range of $H>H_c$ to the initial decrease in the hopping energy (while $H=const.$) and next some increase in that energy can be observed. 

Another result, which we think is worth mentioning, is the switching effect of the pair-cluster magnetization caused by the magnetic field, when the critical value of this field is crossed. This has been illustrated, for instance, in Fig.~\ref{Fig12}. As we see, the effect is most spectacular in the ground state and for the electron concentration $x=1$, which corresponds to the half-filling. The explanation of this phenomenon is strictly quantum. Namely, for $x=(\left< n_a\right>+\left< n_b\right>)/2=1$, the paramagnetic ground state can be realized when $\left< n_{a,{\uparrow}}\right>=\left< n_{a,\downarrow}\right>=1/2$ and $\left< n_{b,\uparrow}\right>=\left< n_{b,\downarrow}\right>=1/2$, i.e., for the case of uniform distribution of two electrons with opposite spins over the whole cluster. On the other hand, the ferromagnetic ground state is realized when $\left< n_{a,\uparrow}\right>=\left< n_{b,\uparrow}\right>=1$ and $\left< n_{a,\downarrow}\right>=\left< n_{b,\downarrow}\right>=0$, i.e., for the case when two electrons with the same spin orientation are separately localized on $a$ and $b$ atoms. This can be understood as a ferromagnetic analogy of the Mott-Hubbard (or Slater) transition \cite{Hirsch, Claveau} for $T=0$, which has been found here for the confined system in the field, where the electrons can be either localized on the separate atoms, or distributed equally (delocalized) over the whole cluster. It cannot be excluded that such an effect, manifested simultaneously as the magnetization switching and being tunable by the electric field (see Figs.~\ref{Fig1} and ~\ref{Fig7}), could find some practical application, for instance, for the magnetic recording or in spintronic devices.

It is worth noticing that some of the results obtained for the Hubbard pair-cluster seem to be generic to the Hubbard model at all system sizes, including infinite systems. For instance, the chemical potential $\mu$ for $x=1$, found in \cite{B-Sz}, amounts to $\mu=U/2$ and is the same as for the infinite system \cite{Pelizzola, Nolting}. Another general property is the electron-hole symmetry with respect to the half-filling case $(x=1)$ \cite{Nolting, Schumann}, which results here in the symmetric form of Figs.~\ref{Fig7} and \ref{Fig12}. The negative, antiferromagnetic correlation function for $T \to 0$ found in this paper (Fig.~\ref{Fig3}) has also been reported both for the infinite systems \cite{Nolting}, and for the finite ($8 \times 8$) 2D lattice \cite{Hirsch}, as well as for the cubic cluster \cite{Schumann}.

The switching effect, occurring in the critical magnetic field $H_c$ and $T \to 0$, can also be expected when the field changes its sign on both atoms forming the pair. This will lead to the transition from paramagnetic to antiferromagnetic (insulating) state, where both electrons with opposite spins are localized on different atoms. A somehow similar transition can take place in the infinite systems, where the external field is not necessary for the phase transition, which instead occurs spontaneously (and might be described using the picture of the internal molecular field of the antiferromagnetic character). As mentioned before, such an effect is known as the Mott-Hubbard transition \cite{Hirsch, Claveau, Sorella,Dang} occurring at zero temperature. From our Fig. 1b it follows that an increase in $U$-parameter at constant field $H$ should facilitate this kind of transition, in accord with the theoretical predictions for 2D and 3D systems \cite{Hirsch, Claveau, Sorella, Hirsch2,Karchev,Kent}. However, the spontaneous magnetic ordering, possible for infinite systems, has not been found here for the pair-cluster, which remains paramagnetic when the external fields are absent. Regarding the influence of the external field, it should be mentioned that the critical magnetic field, when the zero-spin ground state breaks down, has also been found for the cubic cluster \cite{Schumann}, where the exact and rigorous calculations have been performed for the extended Hubbard model.

Finally, it can be concluded, that the method initialized in our previous paper \cite{B-Sz} can be successfully applied to the exact studies of the Hubbard pair-cluster in the external fields with different electron concentration. Further calculations can be based on the higher derivatives of the grand potential, in order to obtain, for instance, the magnetic and electric response functions (susceptibilities) of the system, as well as some magneto- or electrocaloric properties, and thus deserve a separate discussion. 



\end{document}